\documentstyle[preprint,prl,aps]{revtex} 

\begin{document}


\title{Surface-state electron dynamics in noble metals}
\author{P. M. Echenique$^{1,2}$, J. Osma$^{1}$, M. Machado$^{1}$, V. M.
Silkin$^{2}$, E. V. Chulkov$^{1,2}$, and J. M. Pitarke$^{2,3}$}
\address{$^1$ Materialen Fisika Saila, Kimika Fakultatea, Euskal Herriko
Unibertsitatea,\\
1072 Posta kutxatila, 20080 Donostia, Basque Country, Spain\\
$^2$Donostia International Physics Center (DIPC) and Centro Mixto
CSIC-UPV/EHU,\\
Donostia, Basque Country, Spain\\
$^3$ Materia Kondentsatuaren Fisika Saila, Zientzi Fakultatea, 
Euskal Herriko Unibertsitatea,\\ 644 Posta kutxatila, 48080 Bilbo, Basque 
Country,
Spain}

\date\today

\maketitle

\begin{abstract}
Theoretical investigations of surface-state electron dynamics in noble metals
are reported. The dynamically screened interaction is computed, within
many-body theory, by going beyond a free-electron description of the metal
surface. Calculations of the inelastic linewidth of Shockley
surface-state electrons and holes in these materials are also presented. While
the linewidth of excited holes at the surface-state band edge (${\bf
k}_\parallel=0$) is dominated by a two-dimensional decay channel, within
the surface-state band itself, our calculations indicate that major
contributions to the electron-electron interaction of surface-state electrons
above the Fermi level come from the underlying bulk electrons.    
\end{abstract}


\section{Introduction}
Shockley surface states are known to exist near the Fermi
level in the $\Gamma$-L projected bulk band gap of the (111) surfaces of the
noble metals Cu, Ag, an Au\cite{Gartland,Plummer,Himpsel}. The wave functions of
these crystal-induced surface states\cite{Smith,Echenique1} are localized near
the surface and decay into the solid, in contrast to the Bloch waves propagating
into the bulk. Hence, these states form a quasi two-dimensional (2D) electron
gas, which overlaps in energy and space with the three-dimensional (3D)
substrate, and represent a playground for lifetime investigations.

Various techniques have been used to measure the lifetime broadening of
surface-state electrons and holes near the Fermi level. High-resolution
{\it angle-resolved photoemission spectroscopy} (ARPES) has been used to investigate the lifetime
of surface-state
holes\cite{Jensen,Panagio,Theilmann,Matzdorf,Goldmann,Balasubramanian}. In
contrast to this technique, which is restricted to lifetime measurements of
occupied states, it has been demonstrated that {\it scanning tunneling spectroscopy}
(STS) offers the possibility to measure the lifetime of long-living surface states
above and below the Fermi level\cite{Li,Kliewer,Burgi}. The lifetime of excited
holes at the band edge of the partially occupied Shockley surface states on
Ag(111), Cu(111), and Au(111)  was accurately determined by Li {\it et
al}\cite{Li} and by Kliewer {\it et al}\cite{Kliewer} with the use of STS.
Recently, STS has been used to measure lifetimes of hot surface-state and
surface-resonance electrons, as a function of excess energy\cite{Burgi}.

On the (111) surfaces of Cu, Ag, and Au, the Shockley surface state at the
center of the surface Brillouin zone (${\bf k}_\parallel=0$) lies just below the
Fermi level, with $E-E_F=-445$, $-67$, and $-505\,{\rm meV}$, respectively
\cite{Kliewer}, $E$
and $E_F$ representing the surface-state energy and the Fermi level. The
dispersion relation of these states is displayed in Fig. 1, which shows that
they are free-electron-like with effective masses of $0.42$, $0.44$, and
$0.28\,m_e$ ($m_e$ is the free-electron mass) that account for the potential
variation parallel to the surface. This figure clearly shows that the decay of
Shockley surface-state electrons and holes in these materials may proceed either
through the coupling with bulk states (3D channel) or through the coupling,
within the surface-state band itself, with surface states of different wave
vector parallel to the surface (2D channel).

While 3D free-electron calculations of the decay rate of excited holes at the
surface-state band edge (${\bf k}_\parallel=0$) of the noble metals are known to
predict significantly longer lifetimes than those observed with either
ARPES\cite{Jensen,Panagio,Theilmann,Matzdorf,Goldmann,Balasubramanian} or
STS\cite{Li,Kliewer}, recent measurements\cite{Burgi} have shown that, in the case
of surface-state electrons above the Fermi level, the experimental values are
comparable to the calculated lifetimes of bulk free-electrons with the same
energy. That the decay of surface-state holes is dominated by 2D
electron-electron interactions, but screened by the underlying 3D electron
system, has been demonstrated recently\cite{Kliewer,Chulkov}, showing an excellent
agreement with experiment. In this paper, we extend these theoretical
investigations to the case of hot surface-state and surface-resonance electrons
in Cu(111), and show that major contributions to the electron-electron
interaction of surface-state electrons above the Fermi level come from the
underlying bulk electrons. We also present self-consistent calculations of the
screened interaction, and investigate the role that both the partially occupied
2D surface-state band and the underlying 3D electron
gas play in the relaxation mechanism.

The contribution to measured lifetimes of surface-state electrons in the noble 
metals arising from electron-phonon interactions has been discussed in 
\onlinecite{Kliewer}. Here we focus on the investigation of 
the energy-dependent inelastic lifetimes that are due to electron-electron
interactions. 

\section{Theory}

We take an arbitrary Fermi system of interacting electrons and consider an
excited electron or hole interacting with the Fermi sea. The electron-hole
decay rate, i.e., the probability per unit time for the probe particle to scatter
from an initial state $\psi_i({\bf r})$ of energy $E_i$ to some available final
state $\psi_f({\bf r})$ of energy $E_f$ ($|E_f-E_F|<|E_i-E_F|$), by carrying the
Fermi gas from the many-particle ground state to some excited many-particle
state, may be obtained by using the {\it golden rule} of
time-dependent perturbation theory\cite{Shiff}. By keeping terms of first-order
in the screened interaction, one finds (we use atomic units throughout, i.e.,
$e^2=\hbar=m_e=1$):
\begin{eqnarray}\label{eq92}
\tau^{-1}=&&2\sum_f\int{\rm d}{{\bf r}}\int{\rm d}{{\bf
r}'}\,\psi_i^*({\bf r})\,\psi_f^*({\bf r}')\cr\cr
&&\times{\rm Im}\left[-W({\bf r},{\bf r}';|E_i-E_f|)\right]\,\psi_i({\bf
r}')\,\psi_f({\bf r}),
\end{eqnarray}
where $W({\bf r},{\bf r}';\omega)$ is the
frequency-dependent dynamically screened interaction
\begin{eqnarray}
W({\bf r},{\bf r}';\omega)=&&v({\bf r},{\bf r}')+\int d{\bf r}_1\int d{\bf r}_2\,
v({\bf r},{\bf r}_1)\cr\cr
&&\times\chi({\bf r}_1,{\bf r}_2;\omega)\,v({\bf r}_2,{\bf r}'),
\end{eqnarray}
$v({\bf r},{\bf r}')$ and $\chi({\bf r}_1,{\bf r}_2;\omega)$ being the bare
Coulomb interaction and the {\it exact} density-response function of the Fermi
system, respectively.

In the case of a bounded 3D electron gas that is
translationally invariant in the plane of the surface, which we assume to be
normal to the $z$ axis, the probe-particle initial and final states are of the
form
\begin{equation}
\psi_i({\bf r})=\phi_i(z)\,{\rm e}^{i{\bf k}_\parallel\cdot{\bf r}}
\end{equation}
and
\begin{equation}
\psi_f({\bf r})=\phi_f(z)\,{\rm e}^{i({\bf k}_\parallel-{\bf
q}_\parallel)\cdot{\bf r}},
\end{equation}
with energies
\begin{equation}\label{mi}
E_i=\varepsilon_i+{{\bf k}_\parallel^2\over 2m_i}
\end{equation}
and
\begin{equation}\label{mf}
E_f=\varepsilon_f+{({\bf k}_\parallel-{\bf q}_\parallel)^2\over 2m_f},
\end{equation}
where the one-particle wave functions $\phi(z)$ and energies $\varepsilon$
describe motion normal to the surface. Using these wave functions and energies in
Eq. (\ref{eq92}), one easily finds
\begin{eqnarray}\label{b1}
\tau^{-1}=&&2\,\sum_f
\int{{\rm d}{\bf q}_\parallel\over(2\pi)^2}\int{\rm d}{z}\int{\rm d}{z'}
\,\phi_i^*(z)\,\phi_f^*(z')\cr\cr
&&\times{\rm
Im}\left[-W(z,z';{\bf q}_\parallel,|E_i-E_f|)\right]\,\phi_i(z')\,\phi_f(z),
\end{eqnarray}
$W(z,z';{\bf q}_\parallel,\omega)$ being the 2D Fourier transform
of the screened interaction.

\section{Screened interaction}

The main ingredient in the evaluation of electron-hole decay rates in a bounded
3D electron gas is the screened interaction $W(z,z';{\bf
q}_\parallel,\omega)$. When both $z$ and $z'$ are fixed far from the surface,
a few atomic layers within the bulk, there is translational invariance in
the direction normal to the surface, and $W(z,z';{\bf q}_\parallel,\omega)$ can
then be easily obtained from the knowledge of the dielectric function of a
homogeneous electron gas $\epsilon(q,\omega)$:
\begin{equation}\label{bulk}  
W^{bulk}(z,z';{\bf q}_\parallel,\omega)=\int{dq_z\over
2\pi}\,{\rm e}^{i\,q_z(z-z')}v(q)\,\epsilon^{-1}(q,\omega),
\end{equation}
where $v(q)$ represents the three-dimensional Fourier transform of the bare
Coulomb interaction, and $q^2=q_\parallel^2+q_z^2$. In the {\it random-phase
approximation} (RPA), $\epsilon(q,\omega)$ is the Lindhard dielectric
function\cite{Lindhard}.  

Shockley surface-state electrons and holes are found to be very close to the
Fermi level ($|E-E_F|<<E_F$). Hence, as the energy transfer $|E_i-E_f|$ entering
Eq. (\ref{b1}) cannot exceed the value $|E-E_F|$, one can use the low-frequency
form of the Lindhard dielectric function, which yields the energy-loss function
\begin{equation}\label{omega}
{\rm Im}\left[-\epsilon^{-1}(q,\omega)\right]={2\over
q^3}\,\epsilon^{-2}(q,0)\,\omega\,\Theta(2q_F-q),
\end{equation}
where $\Theta(x)$ represents the Heaviside function and $q_F$ is the Fermi
momentum. If one further replaces the static dielectric function $\epsilon(q,0)$
by the {\it Thomas-Fermi} (TF) {\it approximation}, then one finds
\begin{equation}\label{eloss2}
{\rm Im}\left[-\epsilon^{-1}(q,\omega)\right]={2q\over
(q^2+q_{TF}^2)^2}\,\omega\,\Theta(2q_F-q),
\end{equation}
$q_{TF}=\sqrt{4q_F/\pi}$ being the TF momentum.

In the bulk, the maximum magnitude of ${\rm
Im}\left[-W(z,z';{\bf q}_\parallel,\omega)\right]$ occurs at
$z=z'$ and is independent of the actual value of $z$. Fig. 2 shows this quantity
for Cu(111), as obtained from Eq. (\ref{bulk}) with either the Lindhard
dielectric function or the approximated energy-loss function of Eq.
(\ref{eloss2}), versus $q_\parallel$ for fixed values of $\omega$
[$\omega=0.2,0.3,0.4,0.5\,{\rm eV}$]. One clearly sees that for the
momentum and energy transfers of interest (see Fig. 1), the approximated
form of Eq. (\ref{eloss2}) yields a screened interaction which is close to
that obtained with use of the Lindhard dielectric function, small differences
being mainly due to the fact that the static dielectric function
$\epsilon(q,0)$ has been replaced by the TF approximation.
Contributions to the surface-state decay rate coming from the penetration of the
surface-state wave function into the solid are, therefore, expected to be well described
by Eq. (\ref{eloss2}).

Nevertheless, coupling of Shockley surface states with the crystal may also occur,
either through the evanescent tails of bulk states near the surface or within the
surface-state band itself. These contributions to the surface-state decay rate
are both dictated by the screened interaction at $z$ points near the surface,
where the representation of Eq. (\ref{bulk}) is not accurate.

For a realistic description of the screened interaction, we proceed along the
lines reported in \onlinecite{prl}. First of all, we compute the
single-particle eigenfunctions and eigenvalues of a 1D model
potential\cite{ss}. This model potential, which reproduces far outside the
surface the classical image potential, is chosen so as to describe the width and
position of the energy gap at the $\Gamma$ point (${\bf k}_\parallel=0$) and,
also, the binding energies of both the Shockley surface state at $\Gamma$ and
the first ($n=1$) image-potential induced state. We then introduce these
eigenfunctions and eigenvalues into Eq. (\ref{b1}), we also use them to compute
the non-interacting density-response function\cite{Eguiluz}, and finally solve an
integral equation to derive the RPA density-response function $\chi(z,z';{\bf
q}_\parallel,\omega)$ and the screened interaction $W(z,z';{\bf
q}_\parallel,\omega)$.

As the maximum magnitude of ${\rm Im}\left[-W(z,z;{\bf
q}_\parallel,\omega)\right]/\omega$ near the surface still occurs at $z\sim
z'$ (only far from the surface into the vacuum, where the Shockley
surface-state amplitude is negligible, does the maximum of this quantity
occur at $z\neq z'$\cite{German}), we choose $z=z'$ and show ${\rm
Im}\left[-W(z,z;{\bf q}_\parallel,\omega)\right]/\omega$ in Fig. 3,
as a function of $z$ going from the bulk to the vacuum, for
$q_\parallel=0.2\,a_0^{-1}$ ($a_0$ is the Bohr radius) and for various values of
$\omega$. We find that at the surface ${\rm Im}[-W]/\omega$ is enhanced, for the
smallest values of $\omega$, by a factor of 3 relative to the bulk. This is mainly
due to the strong anisotropy at the surface, which enhances the
electron-hole pair creation probability\cite{note}. We also find
that for frequencies smaller than $0.2-0.3\,{\rm eV}$, the strong anisotropy of
the surface does not modify the linear frequency scaling characteristic of the
{\it bulk} screened interaction [see Eq.(\ref{omega})]. However, for the larger
but still small frequencies explored in Fig. 3, the screened interaction at the
surface exhibits an important enhancement due to corrections to the linear
scaling that are not present within the bulk. As a result, for the largest
values of $\omega$ that we have considered, ${\rm Im}[-W]/\omega$ is enhanced at
the surface by a factor as large as 4.

Fig. 4 displays ${\rm Im}\left[-W(z,z;{\bf q}_\parallel,\omega)\right]/\omega$ as
a function of $z$, for $\omega=0.2\,{\rm eV}$ and for various values of
$q_\parallel$. While ${\rm Im}[-W]/\omega$ is enhanced at the
surface, for the smallest values of $q_\parallel$, by a factor of 3 or 4 in the
frequency range $\omega=0.2-0.5\,{\rm eV}$, this enhancement becomes very small
for the largest values of $q_\parallel$ that we have explored. These
calculations indicate that only transitions with $q<0.4\,a_0^{-1}$ are
affected by the strong anisotropy at the surface.

\section{Lifetime broadening}

We compute the lifetime broadening of Shockley
surface states from Eq. (\ref{b1}) with a realistic description
of the RPA screened interaction, as described in the previous section. Although
Eq. (\ref{b1}) has been derived by assuming that our electron system is
translationally invariant in the plane of the surface, we account for the
potential variation parallel to the surface through the introduction of a
realistic effective mass into Eqs. (\ref{mi}) and (\ref{mf}), and also through
the introduction into Eq. (\ref{b1}) of initial and final wave functions that
change along the actual dispersion curve of each state. The effective mass of
bulk states has been chosen to increase from our computed values of 0.31, 0.25, 
and $0.21\,m_e$ 
at the bottom of the gap in Cu(111), Ag(111), and Au(111),
respectively, to the free-electron mass $m_e$ at the bottom of the valence
band. The $z$-dependent initial and final wave functions $\phi_i(z)$ and
$\phi_f(z)$ have been recalculated for each value of ${\bf k}_\parallel$ and
${\bf k}_\parallel-{\bf q}_\parallel$, respectively, as was done at the
$\Gamma$ point, with use of the 1D hamiltonian described in the
previous section.

As the relaxation of Shockley surface-state electrons and holes may proceed
either through the {\it interband} coupling with bulk states (3D channel) or
through the {\it intraband} coupling, within the surface state itself, with
surface states of different wave vector parallel to the surface (2D channel), we
also consider the decay of electrons and holes in 3D and 2D uniform systems.
Introduction of Eq. (\ref{eloss2}) into Eq. (\ref{bulk}) and then Eq.
(\ref{bulk}) into Eq. (\ref{b1}) with both $\phi_i(z)$ and $\phi_f(z)$ replaced
by plane waves, yields\cite{Quinn}
\begin{equation}\label{eq106}
\tau_{3D}^{-1}={\sqrt{\pi q_F}\over 8q_F^2}
\left[\tan^{-1}\sqrt{\pi q_F}+{\sqrt{\pi q_F}\over
1+\pi q_F}\right]{(E-E_F)^2\over k},
\end{equation}
$k$ being the momentum of the excited electron or hole. On the same
level of approximation, the 2D decay rate is given by\cite{Quinn2}, 
\begin{equation}\label{2D}
\tau_{2D}^{-1}={E_F\over 4\pi}\,
\left[-\ln{|E-E_F|\over E_F}+{1\over 2}+\ln{2q_{2D}^{TF}\over
q_F}\right]\left({E-E_F\over
E_F}\right)^2,
\end{equation}
where $q_{2D}^{TF}=2\,m$ is the TF screening wave
vector in 2D, and $m$ is the electron mass.
 
\subsection{Surface-state holes at $\Gamma$}

The partially occupied Shockley surface-state band forms a uniform
two-dimensional electron gas, with the 2D Fermi energy being given by the band-edge
of the surface state, i.e., $E_F$=445, 67, and 505 eV for the (111)
surfaces of Cu, Ag, and Au, respectively. Our full calculation for the {\it
intraband} linewidth of the band-edge surface-state hole on the (111) surfaces
of Cu, Ag, and Au is presented in Fig. 5 by full squares, together with the
linewidth of Eq. (\ref{2D}) with $E=0$ and the electron mass $m$
chosen to be either the free-electron mass (solid line) or the surface-state
effective mass (full circles). We find that calculations based on a pure 2D
electron gas are larger than the actual {\it intraband} contribution to the
linewidth by a factor of $\sim 7$. This large discrepancy is due to the
fact that electron-electron interactions within the actual Shockley surface-state
2D band are strongly screened by the underlying 3D bulk electron system, thereby
reducing the scattering probability.

Separate {\it intraband} and {\it interband} contributions to the linewidth of
Shockley surface-state holes at the $\Gamma$ point of the projected bulk band-gap
of the (111) surfaces of Cu, Ag, and Au are displayed in Table I. The
3D linewidth of {\it bulk} free holes with the same energy, as obtained from Eq.
(\ref{eq106}), is also shown in this table. Differences between our
full {\it interband} calculations and those obtained from Eq. (\ref{eq106})
arise from (a) the enhancement of ${\rm Im}[-W]$ at the surface, which increases
the linewidth, (b) localization of the surface-state wave function in the
direction perpendicular to the surface, and (c) the restriction that only bulk
states with energy lying below the projected band-gap are allowed.
Both localization of the surface-state wave function and the presence of the 
band-gap reduce the linewidth, and therefore tend to compensate the
enhancement of ${\rm Im}[-W]$ at the surface. In the case of Cu(111) this
compensation is nearly complete, thereby yielding an {\it interband} linewidth
that nearly coincides with the 3D linewidth of free holes. However, this is not
necessarily the case for other materials, depending on the surface band
structure.

The impact of the enhanced ${\rm Im}[-W]$ at the surface on both
{\it interband} and {\it intraband} contributions to the Shockley surface-state
hole linewidth at the $\Gamma$ point in Cu(111) is illustrated in Fig. 6. In this
figure, our full calculations of the {\it interband} and {\it intraband}
contributions to the total inelastic linewidth (full circles) are compared to
the results we have also obtained from Eq. (\ref{b1}), but with the
actual screened interaction replaced by that of Eq. (\ref{bulk}) (full
triangles). 3D and 2D free-electron gas calculations, as obtained from Eqs.
(\ref{eq106}) and (\ref{2D}), are represented by open squares. We find that the
impact of the surface anisotropy on ${\rm Im}[-W]$ (difference between full
circles and triangles) is to largely increase both {\it interband} and {\it
intraband} contributions to the linewidth. Fig. 6 clearly shows that the
agreement, in the case of Cu(111), between our interband linewidth (full circle)
and that obtained from Eq. (\ref{eq106}) (open square)  is due to a fortuitous
cancellation between both localization of the surface-state wave function and
the presence of the band-gap, on the one hand, and the enhancement of ${\rm
Im}[-W]$ at the surface, on the other. We also note from this figure that a
large contribution from {\it intraband} transtions, strongly screened by the
underlying 3D bulk electron system, is responsible for the large differences
between the 3D free-electron gas prediction and the experimental results
(represented by open triangles), as discussed in \onlinecite{Kliewer}.

\subsection{Surface-state electrons and holes with ${\bf k}_\parallel\neq 0$}

Our full calculation of the total inelastic linewidth of Shockley surface-state
electrons and holes on Cu(111), as obtained from Eq. (\ref{b1}), is shown in
Fig. 7 (solid line with circles) versus the surface-state energy. Separate {\it
interband} and {\it intraband} contributions to the linewidth are also
represented in this figure, by solid lines with triangles (interband) and
inverted triangles (intraband), and the linewidth of bulk free-electrons
with the same energy, as obtained from Eq. (\ref{eq106}), is represented by a
dotted line. Measurements of the linewidth of excited holes at the surface-state
band edge (${\bf k}_\parallel=0$) are represented by open triangles, and the STS
measurements for surface-state electrons reported by Burgi {\it et
al}\cite{Burgi} (${\bf k}_\parallel\neq 0$) are represented by full squares.

We find that the fortuitous agreement between the actual
{\it interband} linewidth and that derived from Eq. (\ref{eq106}) is present not
only for surface-state holes at the $\Gamma$ point, but also for surface-state
electrons and holes near the Fermi level. Hence, the interband linewidth shows
the $(E-E_F)^{-2}$ energy dependence that is present in the case of a 3D
free-electron gas.

Fig. 7 shows that at the surface-state band edge (${\bf k}_\parallel=0$ the
intraband linewidth represents an $80\%$ of the total linewidth (see also Table
I). Nevertheless, as the surface-sate wave vector parallel to the surface
increases, the surface-state wave function acquires a bulk-like character, with
a larger penetration into the bulk, and the {\it intraband} contribution to the
linewidth decreases very rapidly. This conclusion explains the experimental
observation that while 3D free-electron gas predictions of surface-state hole
lifetimes are too large, they are in the case of surface-state electrons (above
the Fermi level) comparable to measured lifetimes.

Open triangles in Fig. 7 represent the inelastic contribution to measured
linewidths of Shockley surface-state holes at the
$\Gamma$ point, which are to be be compared with our calculated linewidth at
this point (full circle). Since our model, which correctly reproduces the
behaviour of the
$s$-$p$ valence states, does not include screening from $d$-electrons,
differences between measured linewidths (open triangles) and our calculations
(full circles) are expected to be due to the effect of virtual transitions,
giving rise to additional screening by $d$-electrons\cite{Campillo}, as
discussed in \onlinecite{Kliewer}.

The STS measurements represented in Fig. 7 by full squares include both
inelastic and electron-phonon contributions to the surface-state linewidth.
Hence, one must be cautious in the comparison of these measurements with our
calculations. Electron-phonon linewidths of typically $8.0$ in
Cu(111)\cite{phonons}, which are essentially independent of the surface-state
energy, yield theoretical predictions for the total linewidth that are above the
experimental observation, especially for the lowest surface-state energies
explored. This discrepancy between theory and experiment could be partially
compensated by the screening of
$d$-electrons, which as in the case of surface-state holes at the $\Gamma$ point,
reduce the inelastic linewidth.

\section{Summary and conclusions}

We have reported theoretical investigations of surface-state electron and hole
lifetimes in the noble metals Cu, Ag, and Au. We have also presented
self-consistent calculations of the screened interaction, and have
investigated the role that the partially occupied 2D
surface-state band and the underlying 3D electron gas play in
the relaxation mechanism.

We have reached the conclusion that, while the linewidth of
surface-state excited holes at the $\Gamma$ point of the (111) surfaces of the
noble metals is dominated by a 2D decay channel, major contributions to the
electron-electron interaction of surface-state electrons above the Fermi level
come from the underlying bulk electrons. This key dependence of the relative
contribution of {\it intraband} transitions to the total decay rate explains
the experimental observation that while 3D free-electron gas predictions
of surface-state hole lifetimes are too large, they are in the case of
surface-state electrons (above the Fermi level) comparable to measured
lifetimes.  

\acknowledgments

We acknowledge partial support by the University of the Basque Country, the Basque
Hezkuntza, Unibertsitate eta Ikerketa Saila, and the Spanish Ministerio de Educaci\'on
y Cultura and the Max Planck Research Award funds.

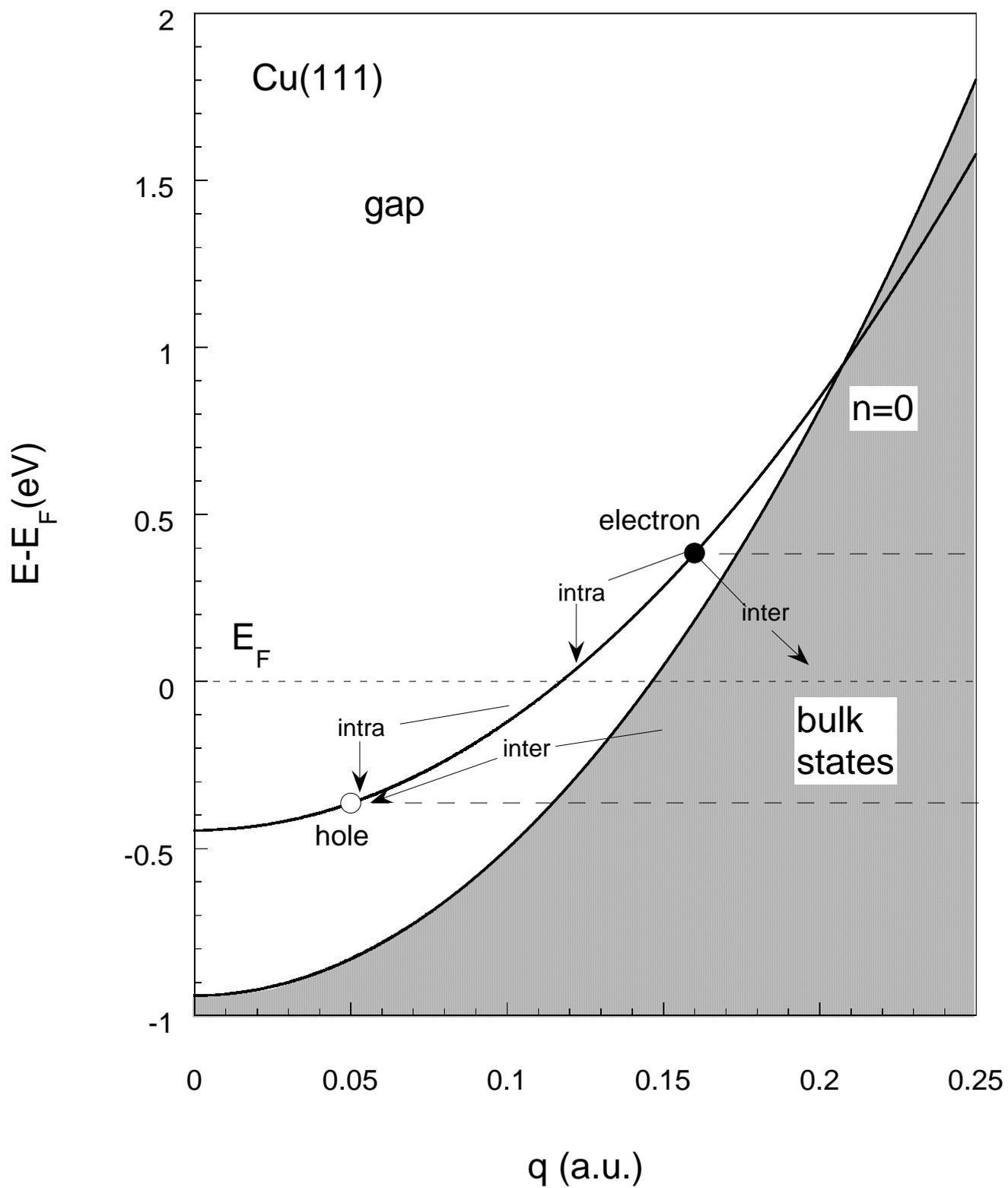
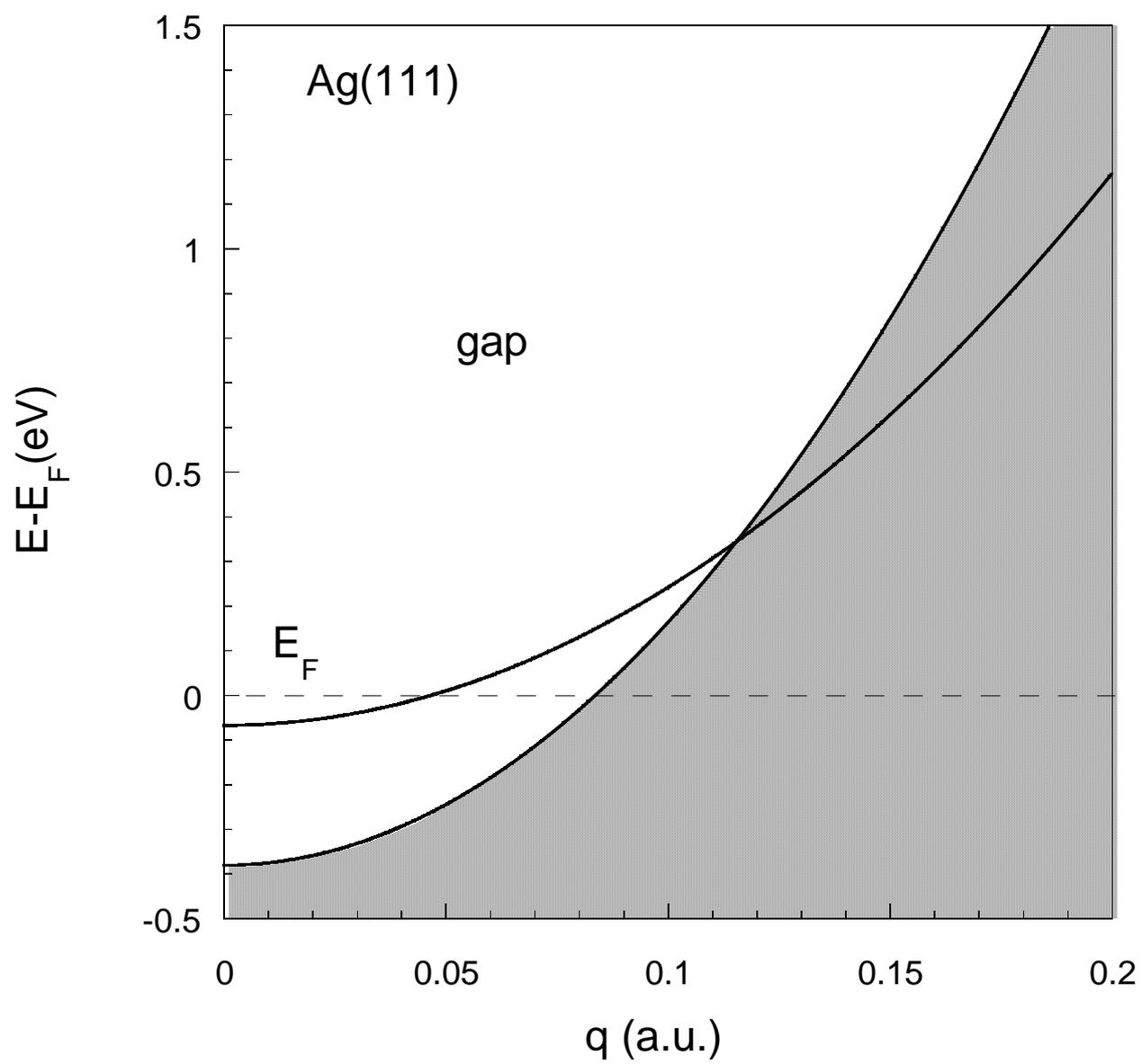
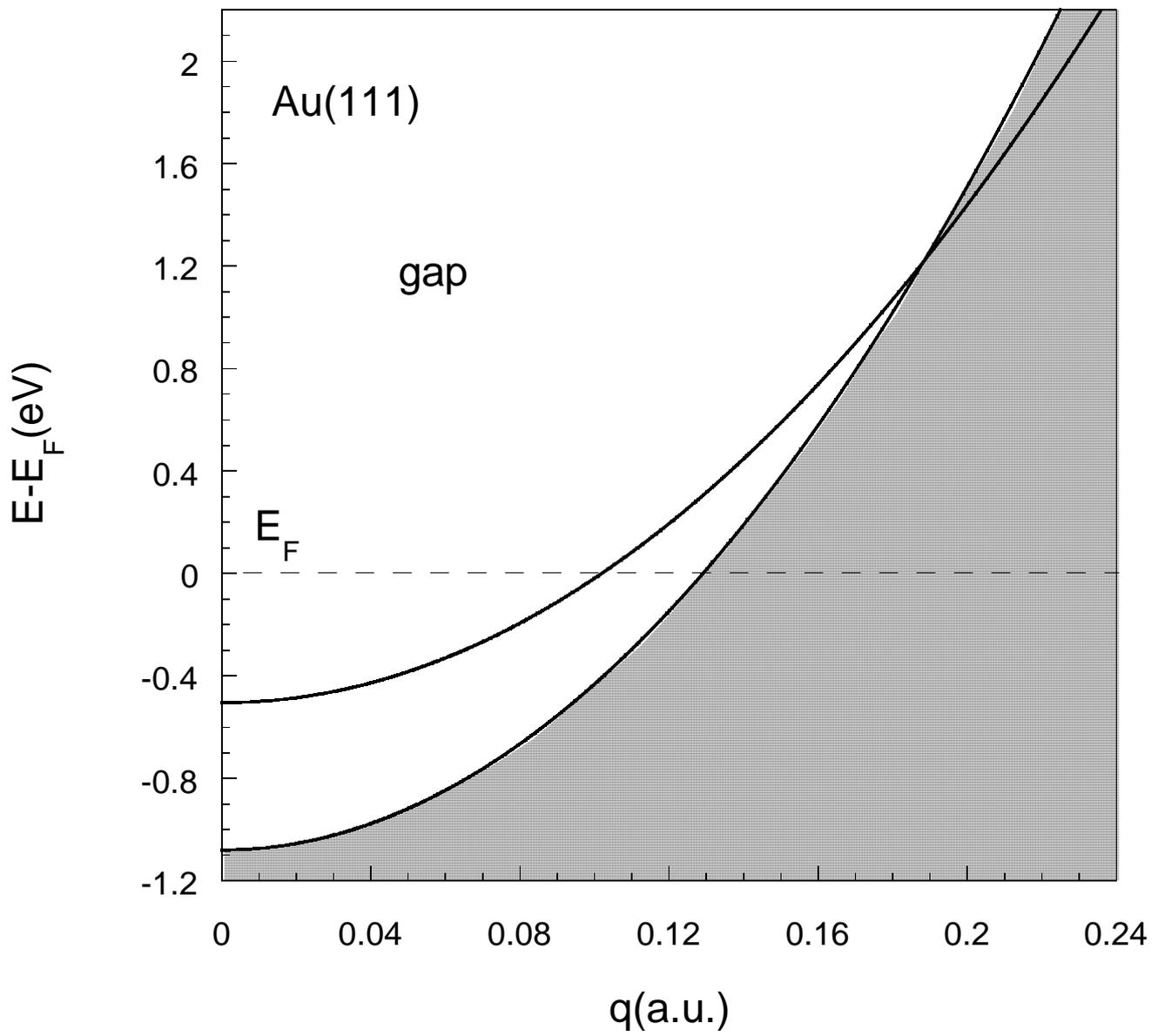
\begin{figure}
\caption{Dispersion of Shockley surface state and bottom of 
projected band-gap on the (111) surfaces of (a) Cu, (b) Ag, and (c) Au. Shaded
areas represent areas outside the band-gap, where bulk states exist. 
Relaxation of Shockley surface-state electrons and holes may proceed either
through interband coupling with bulk states  (3D channel) or through 
intraband coupling, within surface state itself, with surface states of
different wave vector parallel to surface (2D channel).}
\end{figure}

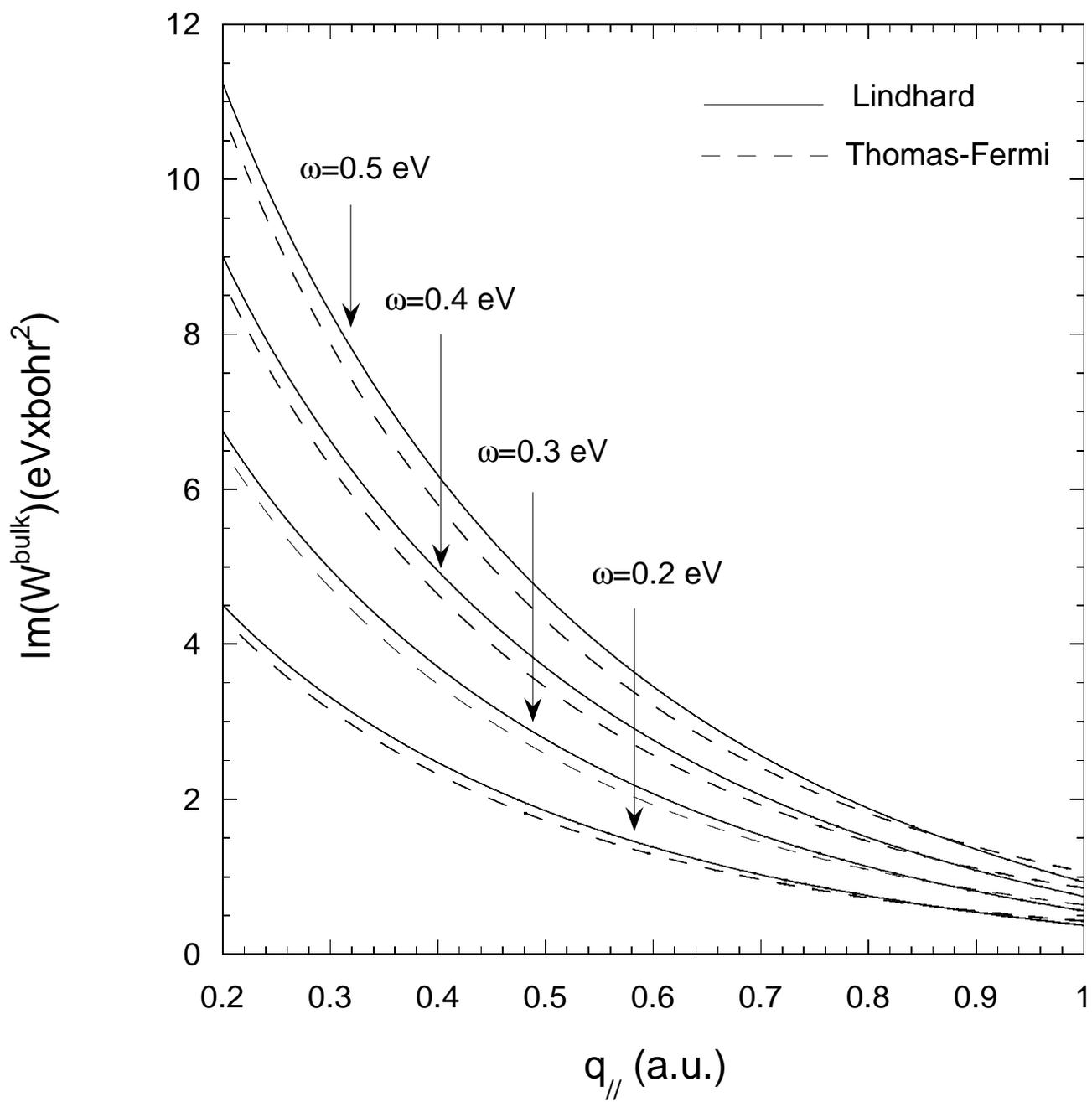
\begin{figure}
\caption{Imaginary part of bulk screened interaction ${\rm
Im}\left[-W^{bulk}(z,z;{\bf q}_\parallel,\omega)\right]$ of Eq. (\ref{bulk}), as
a function of $q_\parallel$ for various values of $\omega$: 0.2, 0.3, 0.4, and
$0.5\,{\rm eV}$. Solid and dashed lines represent the result of introducing
into Eq. (\ref{bulk}) either Lindhard dielectric function (solid lines) or
the approximated energy-loss function of Eq. (\ref{eloss2}) (dashed lines).}
\end{figure}

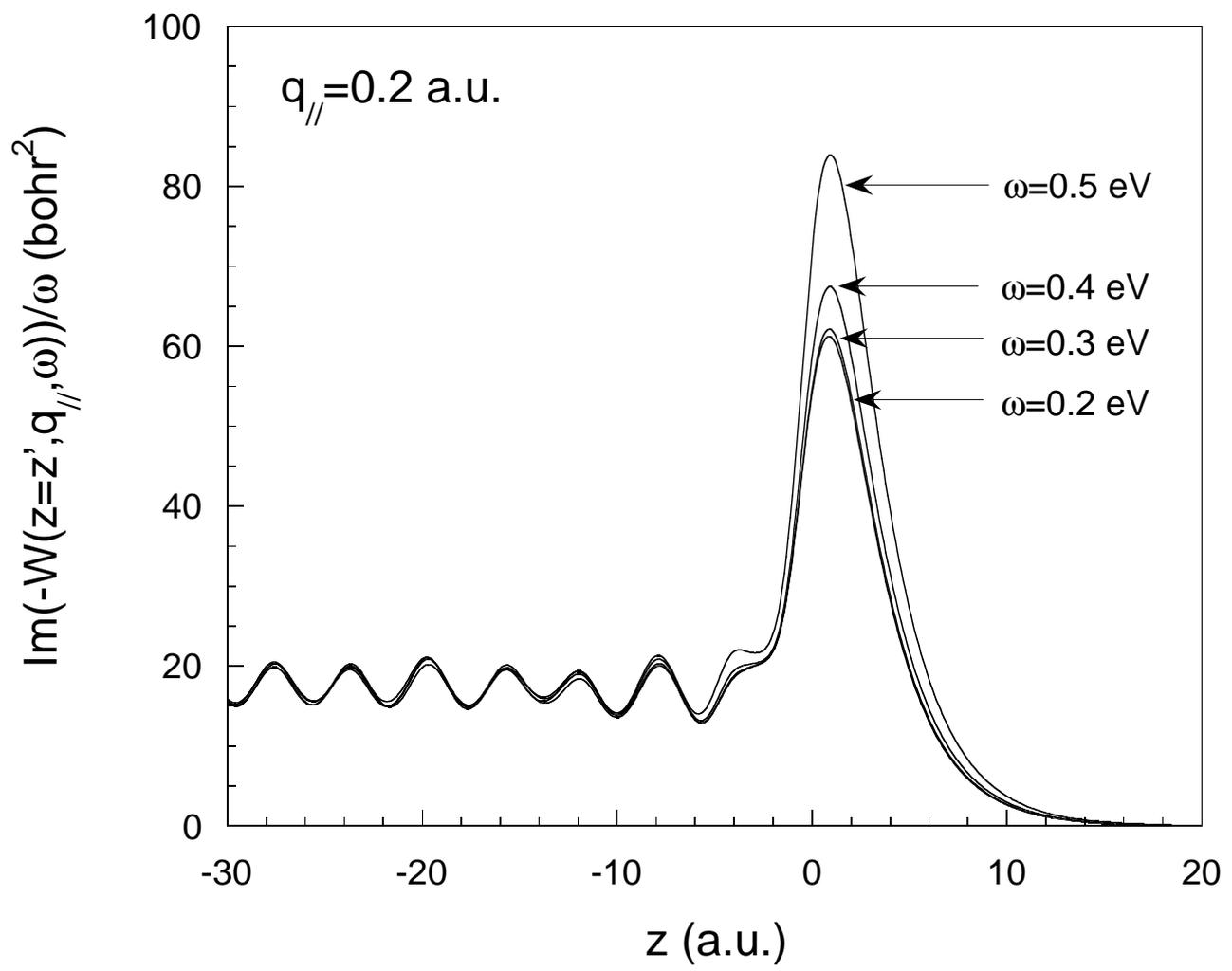
\begin{figure}
\caption{Imaginary part of the scaled RPA screened interaction ${\rm
Im}\left[-W(z,z;{\bf q}_\parallel,\omega)\right]/\omega$ of Eq. (\ref{b1}) ,
versus $z$, for $q_\parallel=0.2\,a_0^{-1}$ ($a_0$
is the Bohr radius) and various values of $\omega$: 0.2, 0.3, 0.4, and
$0.5\,{\rm eV}$.}
\end{figure}

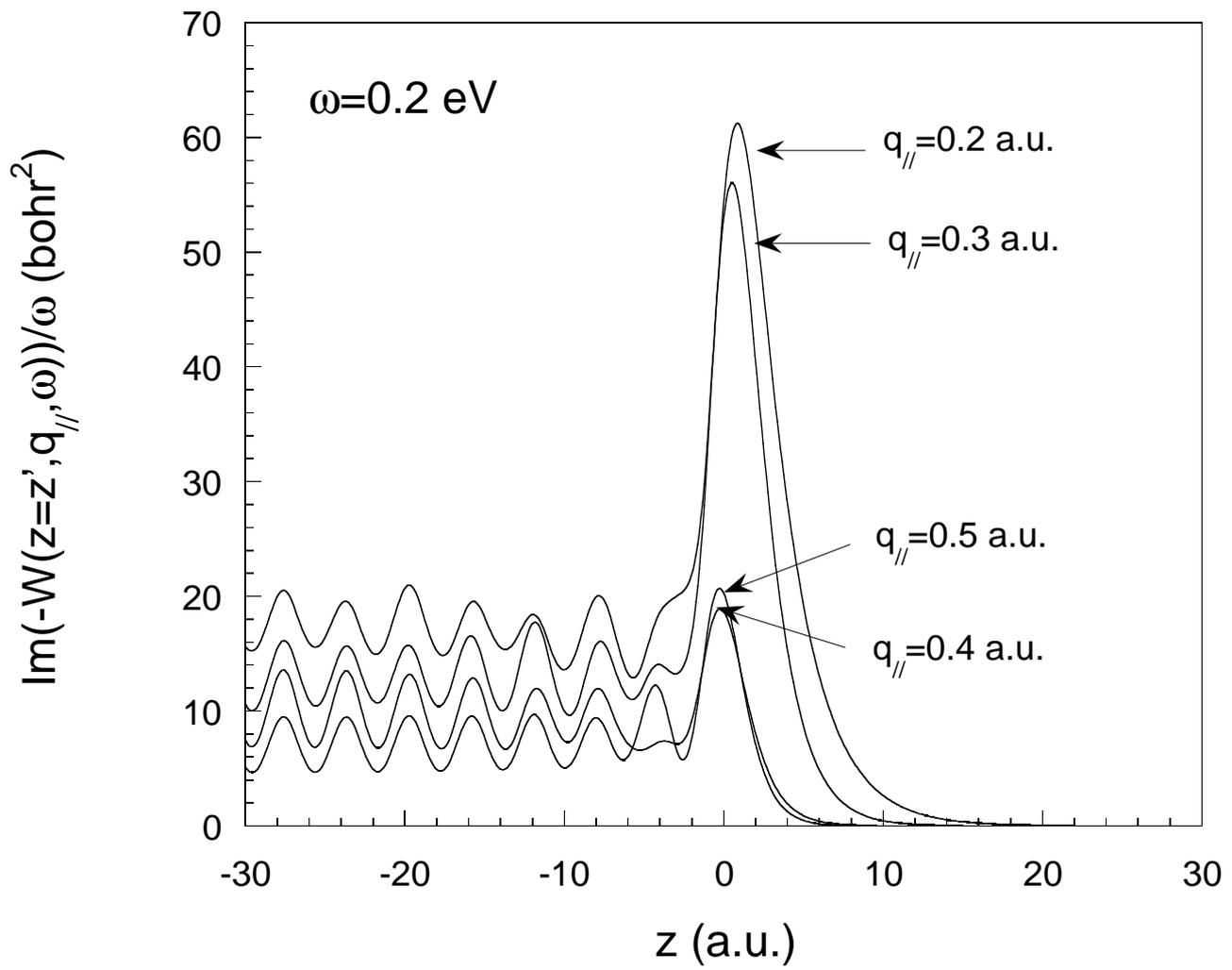
\begin{figure}
\caption{As in Fig. 3, for $\omega=0.2\,{\rm eV}$ and various values of
$q_\parallel$: 0.2, 0.3, 0.4, and $0.5\,a_0^{-1}$.}
\end{figure}

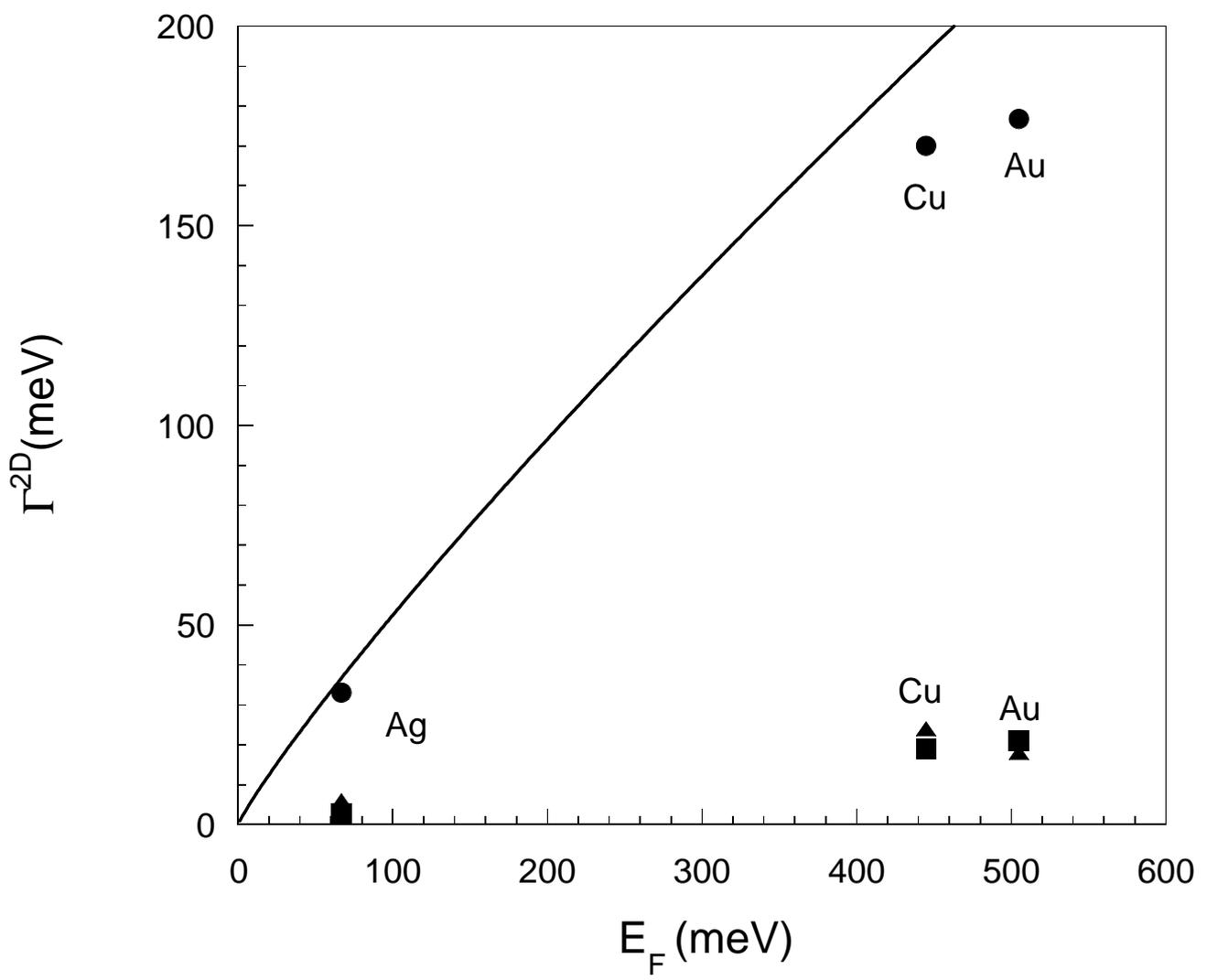
\begin{figure}
\caption{Solid line represents linewidth of holes at the bottom ($E=0$)
of a 2D free-electron gas, as obtained from Eq. (\ref{2D}) with $m=1$, versus
the the 2D Fermi energy. Squares represent our full calculation for the
intraband contribution to the linewidth of band-edge (${\bf k}_\parallel=0$)
Shockley surface-state holes on the noble metals Cu, Ag, and Au. 2D
free-electron gas predictions of Eq. (\ref{2D}) with $E_F=445\,{\rm meV}$ and
$m=0.42$ (Cu), $E_F=67\,{\rm meV}$ and $m=0.44$ (Ag), and $E_F=505\,{\rm meV}$
and $m=0.28$ (Au) are represented by circles. Triangles represent 
experimentally determined inelastic linewidths taken from
\protect\onlinecite{Kliewer}\protect\,(after subtraction of estimated
electron-phonon linewidth of 8, 5.2, and $5.2\,{\rm meV}$ for Cu(111), Ag(111),
and Au(111), respectively.}
\end{figure}

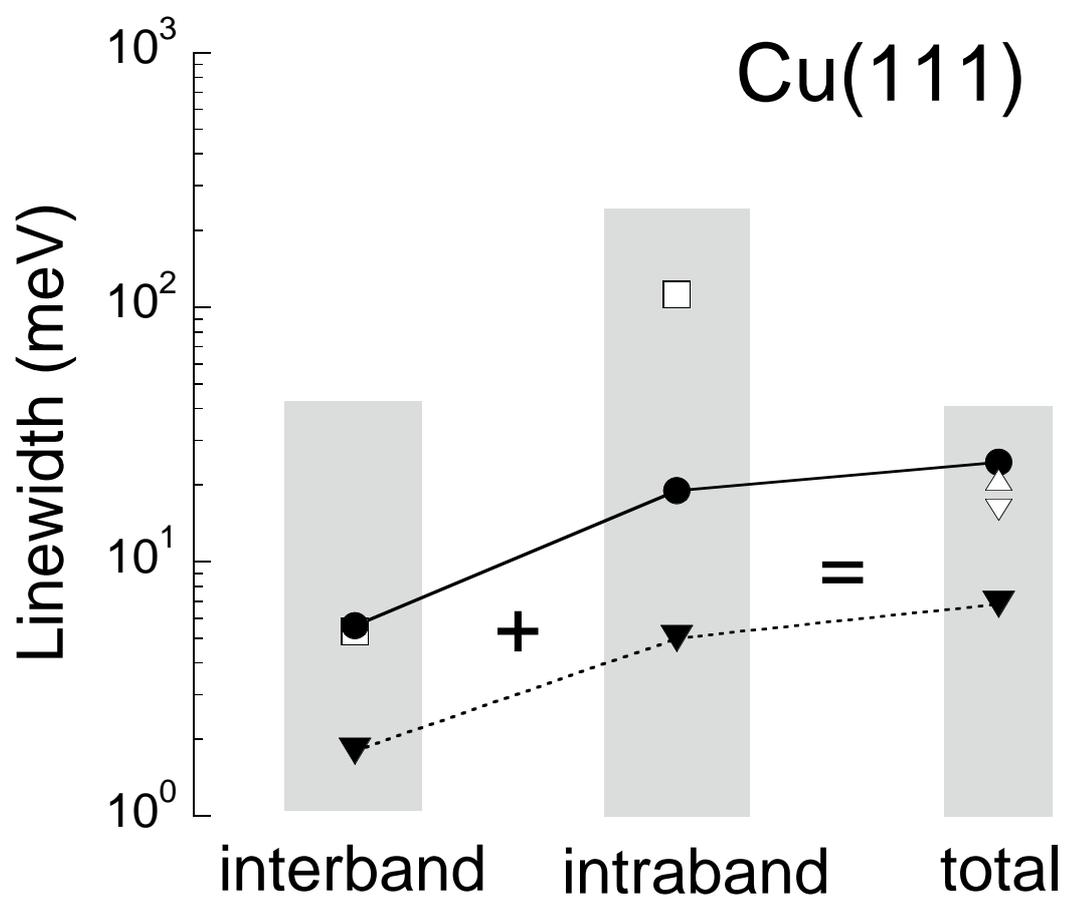
\begin{figure}
\caption{Band-edge (${\bf k}_\parallel=0$) Shockley surface-state hole linewidths
in Cu(111). Full circles: Separate interband and intraband contributions to
total inelastic linewidth, as obtained from Eq. (\ref{b1}) with our full surface
calculation of
$W(z,z';{\bf q},|E_i-E_f|)$. Full triangles: Separate interband and intraband
contributions to the total inelastic linewidth, as obtained from Eq. (\ref{b1})
with $W(z,z';{\bf q},|E_i-E_f|)$ replaced by that of Eq. (\ref{bulk}). Open
squares: 3D and 2D free-electron gas calculations of interband and intraband
linewidths, as obtained from Eqs. (\ref{eq106}) and (\ref{2D}), respectively. 
Experimentally determined inelastic linewidths of
\protect\onlinecite{Theilmann}\protect\, (after extrapolation of ARP
linewidth to zero defect density) and
\protect\onlinecite{Kliewer}\protect\, (after subtraction of estimated
electron-phonon linewidth of $8\,{\rm meV}$) are represented by the open
triangle and the inverted open triangle, respectively.}
\end{figure}

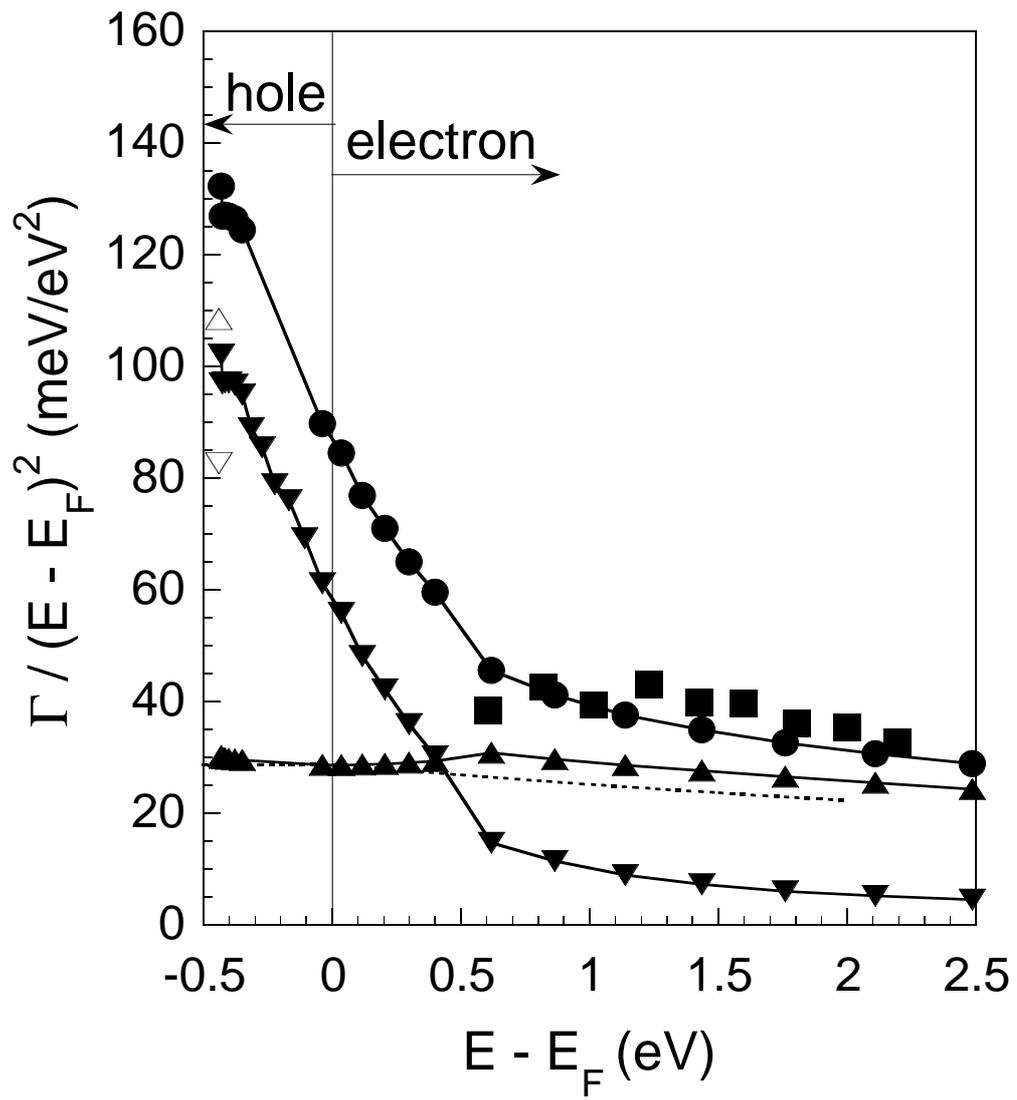
\begin{figure}
\caption{Scaled inelastic linewidth of Eq. (\ref{b1}), $\tau^{-1}/(E-E_F)^2$, of
Shockley surface-state electrons and holes in Cu(111), as a function of 
surface-state energy. Total linewidth is represented by a solid line
with circles. Intraband and interband contributions to the
linewidth are represented by solid lines with triangles
(interband) and inverted triangles (intraband). 3D free-electron gas
prediction of Eq. (\ref{2D}) is represented by a dotted line. Experimentally
determined inelastic linewidths of \protect\onlinecite{Theilmann}\protect\,
(after extrapolation of the ARP linewidth to zero defect density) and
\protect\onlinecite{Kliewer}\protect\, (after subtraction of an estimated
electron-phonon linewidth of $8\,{\rm meV}$) are represented by the open
triangle and the inverted open triangle, respectively, as in Fig. 6. Full
squares represent STS measurements of linewidth (with no subtraction of
electron-phonon contribution) of surface-state and surface-resonance electrons
reported in \protect\onlinecite{Burgi}\protect.}
\end{figure}    

\begin{table}
\caption{Decay rates, in linewidth units (meV), of the Shockley
surface-state hole at the $\Gamma$ point of the noble metals. Decay rates in a 3D
free-electron gas of holes with the energy of the Shockley surface-state at
$\Gamma$ are also displayed.}  
\begin{tabular}{lcccccccc}
Surface&Energy&3D&Inter&Intra&Total\\
\tableline
Cu(111)&-445&5.9&6&19&25\\
Ag(111)&-67&0.18&0.3&2.7&3\\
Au(111)&-505&10&8&21&29\\
\end{tabular}
\label{table1}  
\end{table}

\end{document}